# Building a Casimir Metrology Platform with a Commercial MEMS Accelerometer

Alexander Stange[1], Matthias Imboden[6], Josh Javor[2], Lawrence K. Barrett[1], and David J. Bishop[1,2,3,4,5]

[1]Division of Material Science and Engineering
[2]Department of Mechanical Engineering
[3]Department of Physics
[4]Department of Electrical and Computer Engineering
[5]Department of Biomedical Engineering
Boston University,
Boston, MA 02215
[6]Institute of Microengineering, École Polytechnique Fédérale de Lausanne
2000 Neuchâtel, Switzerland

October 2018

Corresponding author:
Alexander Stange
Photonics Center
8 Saint Mary's Street, Room 609
Boston, MA 02215
267-788-0254, stange@bu.edu

Other emails: mimboden@bu.edu, jjavor@bu.edu, blawrenc@bu.edu, djb1@bu.edu

**Abstract**

The Casimir Effect is a physical manifestation of quantum fluctuations of the electromagnetic vacuum. When two metal plates are placed closely together, typically much less than a micron, the long wavelength modes between them are frozen out, giving rise to a net attractive force between the plates, scaling as $d^{-4}$ (or $d^{-3}$ for a spherical-planar geometry) even when they are not electrically charged. In this paper we observe the Casimir Effect in ambient conditions using a modified capacitive MEMS accelerometer. Using a feedback assisted pick-and-place assembly process we are able to attach various micro-structures onto the post-release MEMS, converting it from an inertial force sensor to a direct force measurement platform with pN resolution. With this system we are able to directly measure the Casimir force between a silver-coated microsphere and gold-coated silicon plate. This device is a step towards leveraging the Casimir Effect for cheap, sensitive, room temperature quantum metrology.

# Introduction

One of the most commonly used technologies enabled by micro-electromechanical systems (MEMS) is the accelerometer.[1,2] These devices are a part of our everyday lives, from sensing the orientation of smartphones (low *g*) to detecting collisions in automobiles in order to deploy airbags (high *g*). Current state-of-the-art low *g* MEMS accelerometers are capable of sensing sub m*g* accelerations with noise densities of around 0.1 mg/Hz$^{1/2}$ or less.[3-5] A typical MEMS accelerometer proof-mass is roughly one microgram, meaning the devices are capable of resolving forces below 1 pN. Such force sensitivity is comparable to the performance of an atomic force microscope (AFM) but is realized on a single mm-scale chip and costs just tens of dollars per device.

In this work we show that by attaching a silver-coated microsphere to its proof-mass, a MEMS accelerometer can directly measure the Casimir force—a quantum fluctuation force that exists between conducting surfaces separated by hundreds of nanometers.[6] First derived by Hendrik Casimir in 1948, the Casimir Effect in its simplest case (two perfectly smooth, perfectly conducting planar surfaces) manifests itself as an attractive force between the two objects which scales as one over separation to the fourth power. The physical origin of this phenomenon is purely quantum mechanical, arising from zero-point fluctuations exerting a net pressure on the conducting surfaces. Because of the small scale of this effect (pN forces at nanometer separations), Casimir force detection nearly always involves a micromechanical system of some kind. Most commonly a modified AFM setup is used, in which a cantilever is adapted to measure forces exerted on it due to Casimir interactions.[7-10] Other work using MEMS has made use of torsional resonators.[11-13] Additionally, devices which can integrate both Casimir surfaces onto a single chip[14] are less prone to low frequency noise and thermal drift due to smaller components and higher mechanical resonant modes but come at the cost of reduced interaction area and limited separation ranges.

The main advantage of a modified accelerometer is the pre-optimized design of the MEMS and the supporting integrated circuitry simplify the device fabrication and apparatus immensely. Despite using an external piezo-mounted plate, we are able to clearly see the Casimir force in ambient conditions. Compared to AFM, the size and cost is superior by orders of magnitude and the linear transduction of an applied force to an electronic output signal is a built-in feature of the accelerometer.

The ability to measure the Casimir Effect with commercial MEMS accelerometers is an exciting prospect because it indicates that this effect could be used as a practical, controllable engineering tool within a MEMS system. For example, it has been suggested that a MEMS oscillator parametrically driven by the Casimir force would exhibit a gain that scales as one over the Casimir cavity size to the fifth power or as applied DC voltage to the tenth power.[15] In addition to providing a means of investigating the Casimir Effect itself, this system could be useful for temperature sensing, AC voltage measurements, low-impedance current measurements, or probing any measurand which could be coupled into a physical movement of the accelerometer proof-mass. Successfully integrating a Casimir cavity into the well-developed, scalable technology of MEMS accelerometers is an important step in realizing Casimir-enabled sensing devices as a practical, room temperature quantum metrology tool.

# Results

## MEMS accelerometer based Casimir cavity

### Micro-gluing onto post-release MEMS

Because the modification involves bonding objects to a post-release MEMS device, great care must be taken in keeping mechanical forces exerted on the freely moving proof-mass to a minimum. We present a technique that allows us to glue microspheres directly to the proof-mass of the accelerometer without compromising the functionality of the MEMS. In this work we use an accelerometer from Analog Devices (ADXL203).

Outlined in figure 1a and 1b is our process which involves depositing ~pL volume droplets of UV curable epoxy using a micro-pipette attached to a piezoelectric actuator onto the proof-mass then dropping a microsphere onto the droplet using a probe tip (contact forces are sufficient to pick up the microsphere). The pipette or probe tip can be moved in plane with a micromanipulator while the Z position is controlled with nanometer precision using the piezoelectric actuator. The advantage of assembling onto a post-release MEMS accelerometer is that we can sense when contact with the proof-mass occurs by actively monitoring the noise on the outputs of the accelerometer (see figure 1a inset). This feedback is what allows us to deposit droplets gently onto the proof-mass without forcing liquid into the release holes or breaking the springs. As can be seen in figure 1c, the proof-mass (shown in blue) provides only a few small areas over which droplets can be placed without interfering with other parts of the MEMS such as the sensing fingers or the springs. In order to attach larger objects, we use one or more spheres (Au-coated solid barium titanate glass) as supports for other objects to be set upon, like legs of a table. Once these "legs" are formed, one can then attach a wide variety of micro-scale objects providing they don't interfere with the MEMS and are able to be picked up and placed gently. For example, it is possible to place a sub-mm neodymium rare-earth magnet on top of the support spheres for high resolution gradient magnetometry.[16] For the device presented in this paper, two 30 μm diameter solid spheres were glued onto the proof-mass as a platform for the rest of the assembly discussed below.

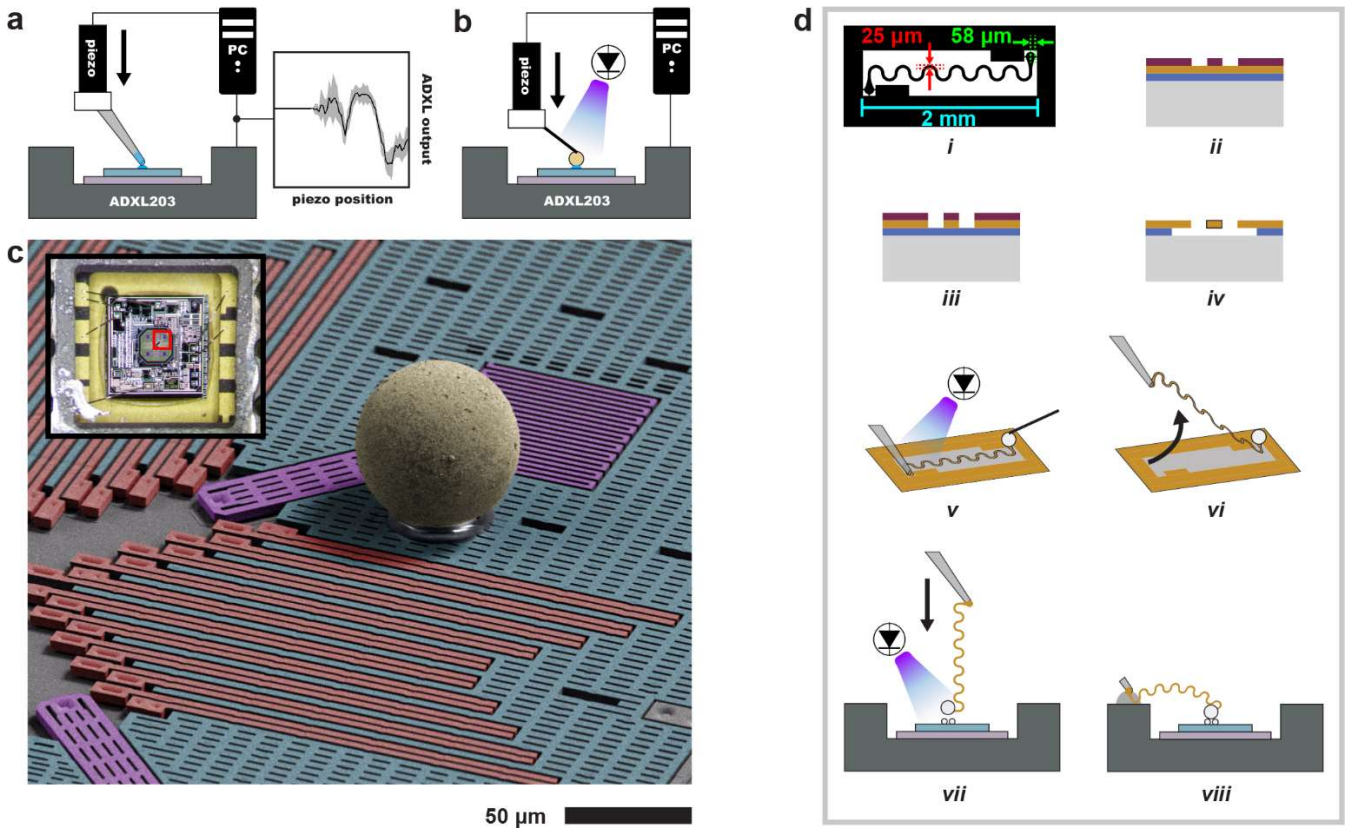

Figure 1: **a.** Schematic of feedback-enabled droplet deposition onto the proof-mass. ADXL output is monitored (inset) while a piezoelectric actuator lowers the micro-pipette (30 μm diameter). Upon contact, surface forces draw out a few pL of epoxy and the pipette is automatically retracted. **b.** Schematic of sphere placement. ADXL output monitored as before while the sphere is lowered into the droplet. Once contact is made, the epoxy is cured by UV exposure. **c.** Colorized SEM image of one quadrant of the MEMS with a microsphere glued to the proof-mass (blue) using the micro-gluing technique. The accelerometer sensing electrodes are shown in red and the springs which anchor the proof-mass to the substrate are shown in purple. **INSET:** top-view optical image of the ADXL203 die inside the package with the lid removed. Highlighted red box indicates the area of the MEMS shown in the SEM image. **d.** Schematic of device assembly steps. The lithography mask (i) for the nano-ribbon wire is designed with a 2 mm nominal length, 25 μm lateral width, and a 58 μm radius circle at each end for attachment. The assembly process involves lowering the Au nano-ribbon wire (with the Ag sphere attached) onto two smaller support spheres which have been previously bonded to the ADXL203 proof-mass using the micro-gluing technique shown in **a**,**b**.

## Casimir force detection

The functional component of our device is a conductive microsphere attached to the proof-mass that forms one half of the Casimir cavity. The sphere is 55 μm in radius and made of hollow borosilicate glass coated with 50 nm of Ag and has a mass of roughly 0.1 μg. It was found that Ag-coated hollow spheres had much lower surface roughness than Au-coated solid spheres (supplementary information S1), so an Ag sphere was used in the Casimir cavity while two smaller Au spheres (about 0.5 μg each) were used as supports. It should be noted that the mass added to the accelerometer proof-mass does not affect the functionality of the device at DC. For dynamic measurements however, the added mass does lower the overall bandwidth of the device.

One requirement in any Casimir device is the ability to control the electric potential on the interacting surfaces. This is due to the presence of residual electrostatic forces, which are caused by trapped charges, adsorbates, and the poly-crystalline nature of the metallic surfaces.[17-19] The latter results in local differences in the work function of the materials (also known as patch potentials) which sum up to a non-zero effective potential difference, even when the materials are electrically connected.[20] This overall residual potential is a common source of error in Casimir force measurements if not controlled for. To do this, a 500 nm thick serpentine ribbon wire connects the surface of the Ag microsphere to an open bonding pad on the accelerometer package. The conductivity of the wire provides a means of controlling the voltage on the sphere and its flexible geometry ensures a low spring constant, thus allowing for minimal restriction of the motion of the proof-mass and the restoring force of the polysilicon springs. The effect that the wire has on the overall effective spring constant of the system (and therefore the force sensitivity) is minimal and easily accounted for as the post-modification force sensitivity is re-calibrated by using known electrostatic forces between the sphere and the plate. The process of assembling this device is shown in schematically in figure 1d and discussed further in Methods.

The second half of the Casimir cavity is formed from a Au-coated plate mounted on a linear piezoelectric actuator. For the device used in this paper, the plate approaches the sphere along the X axis of the accelerometer. The entire setup is mounted on an optical breadboard and contained in a temperature controlled enclosure on top of an active vibration isolation table. A schematic of the device and apparatus can be seen in figure 2c.

## Device performance

### Electrostatic characterization of residual potential and force calibration

Electrostatic forces are used to measure the residual potential difference, $V_0$, between the sphere and the plate and to calibrate the force sensitivity, $\gamma$, which relates the accelerometer output voltage, $S$, to the applied force according to $F = \gamma S$. In figure 2d we plot the voltage output of the accelerometer in the X direction, $S_X$, as we vary the potential applied between the grounded plate and the microsphere ($V_{bias}$) at different separations between 200 nm and 1 µm where electrostatic forces are much larger than the Casimir force. The minima of these voltage sweeps indicate the bias which cancels residual potential between the metal surfaces and the curvature of the sweeps give a calibration coefficient between accelerometer output and force which can then be used for Casimir force measurements (see further discussion on this approach in Methods).

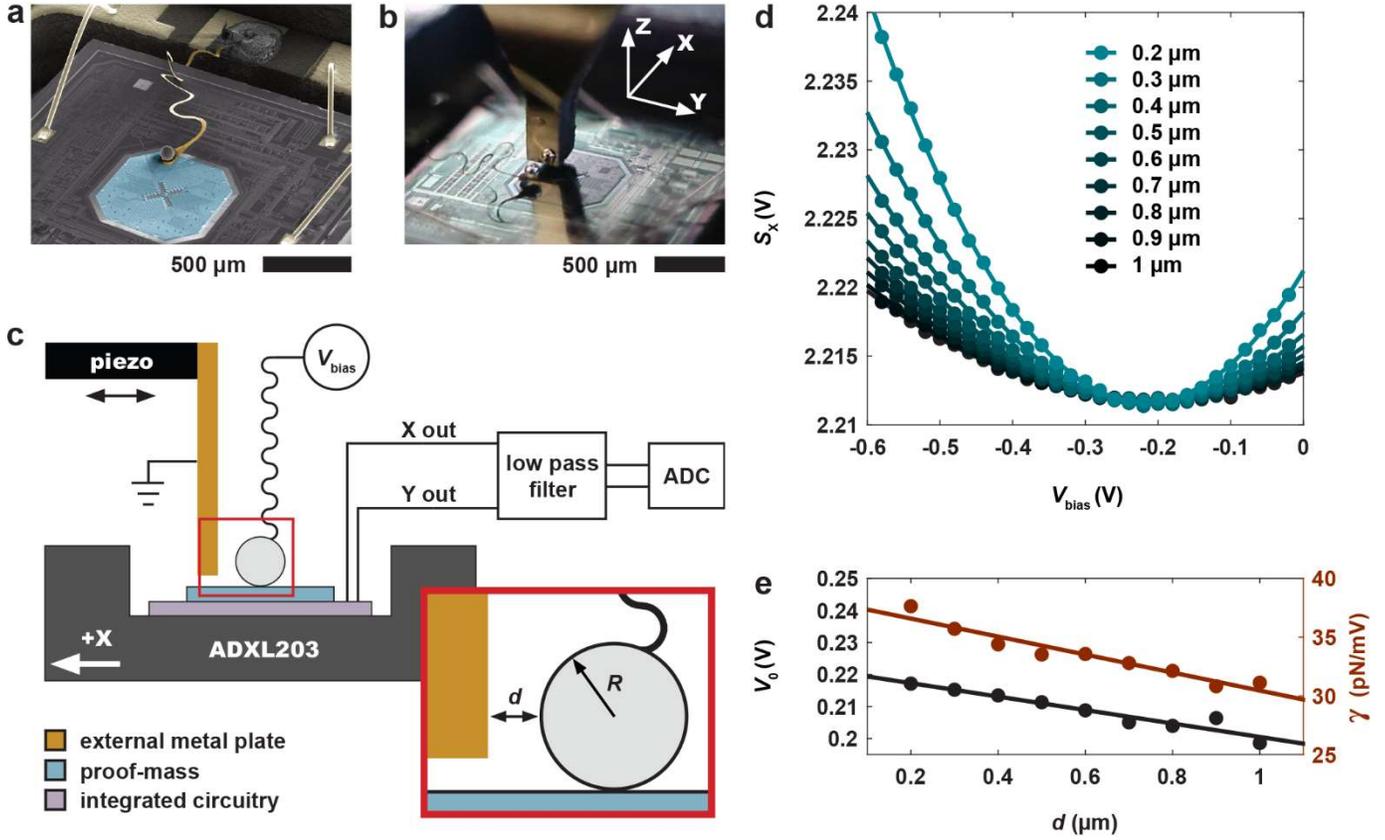

Figure 2: **a.** SEM image of an example of a fully modified accelerometer (not the device used in this work). **b.** Optical image of the modified accelerometer used to collect the data presented in this paper. Also pictured is the external Au-coated plate mounted on a piezoelectric actuator (out of frame). **c.** Schematic of full setup. The accelerometer X and Y outputs are fed through an 8 pole low-pass filter with a 3 Hz cutoff to isolate the desired DC signal and then read by a 16-bit ADC. The Casimir force acts along the X direction for this particular device. **INSET**: Diagram of Casimir cavity geometry showing sphere plate separation ($d$) and sphere radius ($R$). For simplicity, the two support spheres are not pictured. In reality, the Ag sphere is sitting 20 – 30 μm above the proof-mass. **d.** Accelerometer signal data as $V_{bias}$ is varied at different separations. Circles are measured data and the solid lines are second order polynomial fits to the data. **e.** $V_0$ and $\gamma$ versus separation. These values are computed from the minima and curvature of parabolas fit to data in figure **d**.

Both the residual potential and the force sensitivity of the device appear to be functions of separation and are approximately linear. Over the full 800 nm scan range, it is observed that $V_0$ varies by 9% and $\gamma$ varies by 20%, with average values of 0.21 V and 33.5 pN/mV respectively.

**Casimir force measurement**

The data in figure 3 is a measurement of the force applied to the accelerometer proof-mass along its X axis as function of separation between the Ag-coated microsphere (which is attached directly to the proof-mass) and an external Au-coated plate. Electrostatic contributions have been minimized according to the methodology discussed in the following sections. The red and black data points are the same set of data fit to either the ideal Casimir force theory (solid red line) given by equation 8 or the corrected Casimir force theory (solid black line) given by equation 12 with only $x_s$ (i.e. where separation $d = 0$) as a free parameter. According to the ideal fit, the last measured data point at 635.5 pN is 65 nm away from $d = 0$. According to the corrected Casimir fit, the last measured data point is 63 nm away from $d = 0$.

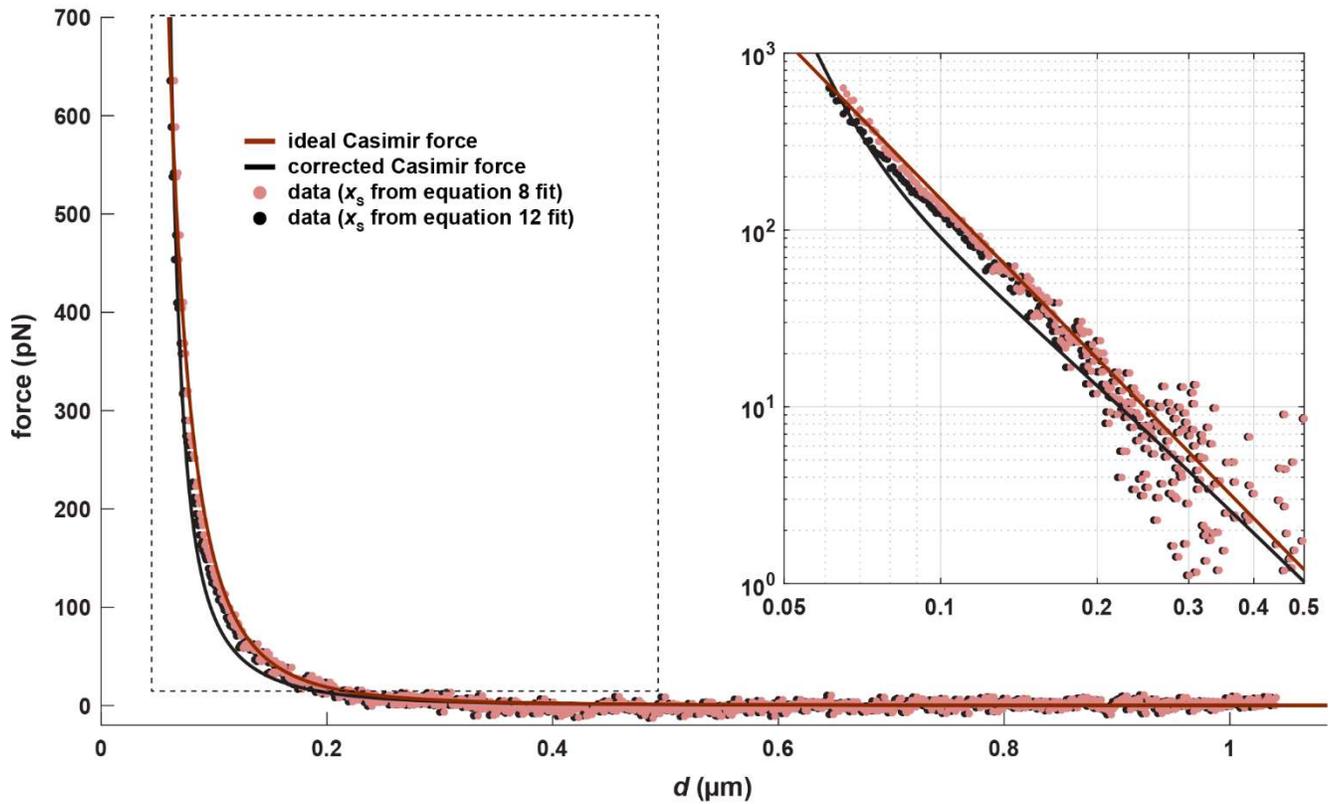

Figure 3: Casimir force measurements compared with ideal theory equation 8 (red) and theory for real metals equation 12 (black). Inset shows the highlighted section of the data in log-log scale for better comparison between data and theory at small separations. The two sets of data are identical but shifted by 2 nm along the abscissa because of the different values of $x_s$ returned from the fits to equations 8 and 12.

## Discussion

### Distance dependence of residual potential and sensitivity

The separation dependence of the residual potential is a well-known occurrence[21-23] as the potential measured is an effective sum of the contributions from different regions across the surfaces of each metal. As the separation changes, these contributions will sum differently due to the inverse square dependence of the electrostatic force. The linear dependence of the force sensitivity of the device is due to the interaction of the grounded Si plate with the fringe fields of the interdigitated capacitor fingers of the accelerometer. The capacitive sensing of the accelerometer relies on a small AC signal applied between the fixed fingers and the movable fingers. Because the plate is held at the same ground as the accelerometer it will deflect fringe field lines from this applied voltage and result in an out of plane force exerted on the grounded fingers.[24,25] As the plate's position is varied it will overlap with more fringe fields, thus exerting a larger out of plane force and decreasing the sensitivity. This effect is more prominent the closer the plate is to the fingers (see supplementary information S2). Therefore, there is an ideal range of plate heights at which the experiment can be performed, where the fringe field interactions are minimized while also ensuring adequate area of interaction between the side of the sphere and the plate.

## Casimir force comparison with theory

It is interesting that the measured data is modeled more accurately by the ideal Casimir theory (root mean squared error of 7.4 pN) compared to the corrected theory (root mean squared error of 10.5 pN). While the ideal theory overestimates the measured force in the 80 nm to 200 nm range the corrected theory underestimates it by far greater. Considering random measurement errors, the linear fits shown in figure 2e returned root mean square errors to the data of 1.8 mV and 0.6 pN/mV for $V_0$ and $\gamma$ respectively. If we assume a mis-calculated $V_0$ off by three standard deviations (5.4 mV) at the closest point of approach of 60 nm, we would be introducing an unwanted additional electrostatic force of 0.74 pN, which is just 0.1% of the Casimir force measured at that point. This value is also much less than the random error in the force measurement due to uncertainty in the product $\gamma S_x$ (whose combined errors propagate to as high as 60 pN at the closest measured point). It is therefore likely that the discrepancy between measurement and theory in this distance range are not due to imperfectly cancelled electrostatic forces, as the Casimir force is the dominant interaction.

The reason for this discrepancy is most likely due to the assumptions of the geometry of the cavity, both at the microscale (i.e. sphere and plate shapes and arrangement) and at the nanoscale (surface roughness). The sphere radius, $R$, is used in both the electrostatic calibration analysis as well as the theoretical Casimir fits. In both cases, we use $R$ in the framework of the Proximity Force Approximation (PFA) which assumes a perfectly spherical surface and an infinite plane in the $d \ll R$ limit.[26] Because of this, both the calibration factor and the Casimir force are proportional to $R$, so any errors in the value of $R$ (which was measured optically) do not affect the fitting. However, if the sphere is not perfectly spherical, then we would expect both the electrostatic force and the Casimir force to scale differently depending on the exact shape.[27] Additionally, due to physical constraints of our setup, the plate is limited to extend only ~80-90 µm below the central plane of the sphere because of the proof-mass. Because the PFA assumes an infinite plane, this asymmetry may result in a systematically overestimated force sensitivity of the device.

The surface roughness also becomes very important in Casimir interactions at separations less than ~100 nm.[28,29] The AFM scans taken were on separate samples which went through the same coating processes as the sphere and plate used in this device. While this may be useful for capturing average roughness values, it is not specific to this exact cavity, which may have extreme asperities that cause deviation from the expected scaling below 100 nm separations. To improve both of these geometry related systematic uncertainties, characterization of individual spheres would need to be done prior to subsequent assembly.

Nevertheless, our results show that an Ag surface and an Au surface in our atmospheric MEMS system exhibit an interaction which can be described quite well by the ideal Casimir force model. This is a promising finding as we move forward with Casimir-enabled sensing devices, such as that described by Imboden *et al.* in ref. 15. The results presented here imply that approximating this interaction as a simple inverse cubic relation is quite sufficient for further analysis and modeling.

## Conclusions

The purpose of this work was two-fold: First, we have successfully integrated a Casimir cavity onto a MEMS system which is capable of resolving pN forces. This system provides the experimenter with a customizable apparatus for investigating the Casimir Effect with non-trivial geometries, materials, or surface morphologies such as nanostructures or chiral metamaterials.[30-34] Additionally, this work is an important stepping stone in our goal to leverage the extraordinary distance dependence of the Casimir force to provide an enhancement in the sensitivity of the device through parametric techniques. Second, we have shown that it is possible to perform highly sensitive quantum metrology in ambient conditions with off-the-shelf consumer accelerometers, which are widely available and very inexpensive. MEMS accelerometers can be used as a novel tool for experimenters—a literal platform capable of performing a variety of interesting micro and nano-scale low-force experiments. Using the feedback-assisted micro-gluing process we have developed, one can re-purpose the accelerometer to transduce any measurand that can be coupled into a displacement of the proof-mass.

# Materials and methods

## MEMS accelerometer

The fundamental building block of our Casimir measurement system is a MEMS accelerometer from Analog Devices (ADXL203)—a two-axis capacitive accelerometer with analog voltage outputs for the in-plane X and Y directions with a sensitivity of ~1 V/$g$ along both axes.[5] The polysilicon MEMS consists of an octagonal proof-mass that is anchored to the substrate via four serpentine springs with a total effective spring constant of roughly 1 N/m. The proof-mass/spring system has a fundamental resonant mode at 5.4 kHz with a quality factor of 10 in air and 1000 in vacuum (~1 × $10^{-4}$ Torr). The proof-mass also has four sets of finger electrodes, forming an interdigitated differential capacitor with another set of fingers anchored to the substrate. Any force applied to the proof-mass moves it, thus changing this capacitance. The proof-mass is surrounded by integrated circuitry on the same chip that demodulates the signals from the differential capacitance measurements and rectifies them into two independent output voltages (nominally 2.5 V with zero applied acceleration) that are proportional to the position of the proof-mass in the X and Y directions. Because the springs obey Hooke's Law, the outputs are linearly proportional to the forces on the proof-mass.  This platform has been optimized with its sensing circuit integrated with the MEMS process to produce a very low noise system which can resolve forces of ~1 pN applied directly to the proof-mass.  The package is easily opened with a straight edge razor blade.  After the lid is removed, the MEMS proof-mass and electronics can be seen as shown in figure 1c inset.

## Device Assembly

To assemble this device, the following process (outlined in figure 1d) was developed: 1) A serpentine wire with circular ends is fabricated by lithographically patterning and etching a 500 nm thick layer of Au on top of a 2 μm layer of thermally grown oxide. The wire is then released by removing the oxide in hydrofluoric acid (i-iv). 2) A single 55 μm radius hollow borosilicate glass sphere coated with 50 nm of Ag (from Cospheric Technologies) which will function as one half of the Casimir cavity is glued to one end of the serpentine wire using Ag epoxy (Lake Shore Cryotronics, cured by heating at 80°C for 3 hours). This provides electrical connection between the wire and the surface of the sphere. 3) The wire is picked up by the end opposite to the sphere by gluing it to the end of a glass micro-pipette using UV curable epoxy (v,vi). 4) The wire/sphere is lifted off of the substrate, dipped in UV curable epoxy (Norland Optical NOA81), and positioned over the support spheres. 5) The wire/sphere is lowered using a piezoelectric actuator. Contact is determined by monitoring the noise on the X and Y outputs of the accelerometer, as with the gluing process shown in figure 1a inset. When contact is made, the epoxy is cured by UV exposure with a 365 nm LED source (vii). 6) The other end of the wire which is glued to the micro-pipette is brought over to a pad on the ceramic package and glued down using Ag epoxy. 7) Finally, the wire is released by breaking off the end of the micro-pipette (viii). An example of a fully assembled device can be seen in figure 2a. The device used for the results in this paper (along with the external Au-coated plate) can be seen in the optical image in figure 2b.

The plate is fabricated from a 300 μm thick silicon wafer, etched into a tapered shape using photolithography and DRIE, then coated with 10 nm Cr adhesion layer followed by 150 nm of Au using electron beam evaporation.

## Apparatus and experimental details

The modified ADXL203 is mounted on an XY translation stage attached to an optical breadboard. On the same breadboard is another XYZ stage on which the plate is mounted. This stage has its Z position controlled by a Newport Picomotor stick-slip piezoelectric actuator. An additional Newport NPC3SG piezoelectric stack actuator controls the fine position of the plate in the X direction. The apparatus is contained inside a polystyrene foam container along with a 1000 kΩ power resistor and resistance temperature detector for PID controlled temperature with a 28°C setpoint. Due to building heaters cycling on and off, the maximum temperature variations inside the enclosure over long periods of time are ~12 m°C however for shorter time periods (two hours or less) the temperature can be held to within ~3 m°C (see supplementary information S3). The container is set upon an active vibration isolation table (Herzan TS-140).

The potential between the sphere and plate is controlled by grounding the plate and applying a voltage on the sphere through the nano-ribbon wire using a 1 mV resolution power supply (Keithley Instruments). Due to a proprietary polymer coating on the ADXL203 MEMS and the insulating UV curable epoxy, there is no electrical connection between the sphere and the proof-mass.

The accelerometer output is fed directly into a low-pass filter (Stanford Research Systems SR650) with a 3 Hz cutoff frequency and unity gain. The filter output is then sampled by an ADC (National Instruments NIDAQ).

## Measurement and calibration theory and methods

### Electrostatic force

As discussed previously, electrostatic forces are present between the sphere and plate metal surfaces, even when the two metals are shorted together. By applying a voltage equal and opposite to the residual potential, this unwanted electrostatic effect can by minimized. Additionally, by applying known electrostatic forces between the plate and the sphere, the force sensitivity of the output can be calibrated.

The forces acting on the sphere are assumed to be only due to electrostatic and Casimir interactions. For the following equations we define the separation: $d = x_p - x_s$ where $x_p$ is the absolute position of the plate and $x_s$ is the absolute position of the sphere. Assuming a simple electrostatic model, we can write:

$$F(d, V_{bias}) = \frac{\varepsilon_0 \pi R (V_0 + V_{bias})^2}{d} + F_{Casimir}(d) \tag{1}$$

Where $\varepsilon_0$ is the permittivity of free space, $V_0$ is the residual potential, $V_{bias}$ is the applied DC voltage between the sphere and plate, and $R$ is the radius of the sphere. Equation 1 uses the PFA for a sphere-plate geometry which assumes $d \ll R$. At large separations (typically > 200 nm) and large applied voltages ($V_0 + V_{bias}$ > 100 mV), the Casimir force term is negligible, and the force scales as $V_{bias}^2$. The residual potential can be measured by sweeping the bias voltage and finding the value of $V_{bias}$ at which the force is minimized. At this minimum, the applied bias is equal and opposite to the residual potential.

The accelerometer outputs an analog voltage, so to get a measurement in units of force, a calibration must be performed. Because of the linear response of the device, the output signal, $S$, is proportional to the force applied on the proof-mass by a constant, $\gamma$:

$$F(d, V_{bias}) = \gamma S(d, V_{bias}) \tag{2}$$

For large separations and voltages, we can ignore the Casimir term in equation 1 and can now write:

$$S(d, V_{bias}) = \frac{1}{\gamma} \frac{\varepsilon_0 \pi R (V_0 + V_{bias})^2}{d} \tag{3}$$

We can then write $\gamma$ in terms of the curvature of the signal with respect to $V_{bias}$:

$$\gamma = \frac{2\pi\varepsilon_0 R}{d}\left(\frac{\partial^2 S}{\partial V_{bias}^2}\right)^{-1} \quad (4)$$

From these relations, the residual potential and force sensitivity can be measured by fitting the raw signal data, $S$, to the function $S = c_1 V_{bias}^2 + c_2 V_{bias} + c_3$. Using the returned fitting parameters we can calculate:

$$V_0 = \frac{c_2}{2c_1} \quad (5)$$

$$\gamma = \frac{\varepsilon_0 \pi R}{c_1 d} \quad (6)$$

provided we are in a region where the electrostatic force is dominant over the Casimir force.

The procedure for a single measurement is as follows—first the plate is stepped towards the sphere until the accelerometer senses contact. The plate is retracted by 1 µm and $V_{bias}$ is swept, tracing out a parabola as shown in figure 2d, according to equation 3. More sweeps are taken as the plate is moved closer by steps of 100 nm. Every data point is the average of 50,000 samples taken in 0.5 seconds by the ADC with standard deviations between 0.6 and 0.7 mV. The whole electrostatic measurement takes 3.5 minutes. Over this period of time, thermal drift is negligible (see supplementary information S3). Fitting a second order polynomial to these data sets provides measurements of the residual potential, $V_0$, as well as the sensitivity, $\gamma$, according to equations 5 and 6. These values are plotted in figure 2e.

**Casimir force**

For ideal conditions (absolute zero temperature and perfectly smooth infinitely conducting surfaces) the Casimir force between two plates of area $A$ is given by ref. 6 as:

$$F_C^{0,PP}(d) = \frac{\hbar c \pi^2 A}{240 d^4} \tag{7}$$

Where $\hbar$ is Planck's reduced constant, $c$ is the speed of light in vacuum, and $A$ is the overlap area between the surfaces. Using PFA for a sphere-plate geometry, this becomes:

$$F_C^0(d) = \frac{\hbar c \pi^3 R}{360 d^3} \tag{8}$$

provided $d \ll R$, as in the electrostatic case. In addition to this ideal case, we consider a corrected model from Geyer *et al.* which accounts for more realistic physical effects such as non-zero temperature and the finite conductivity of both metallic surfaces.[35] In this model, a perturbation expansion in powers of the relative penetration depths of electromagnetic oscillations into each metal (using a plasma model) provides a corrected equation for the Casimir force given as:

$$\begin{aligned}F_C^P(d) = F_C^0(d)\Bigg[&1 + \frac{45\zeta(3)}{\pi^3 t^3} - \frac{1}{t^4} - 2\frac{\delta}{d}\left(2 - \frac{45\zeta(3)}{\pi^3 t^3} + \frac{2}{t^4}\right) \\ &+ \frac{72}{5}\frac{\delta^2}{d^2} - \frac{320}{7}\frac{\delta^3}{d^3}\left(1 - \frac{2\pi^2}{105}(1-3\kappa)\right) + \frac{400}{3}\frac{\delta^4}{d^4}\left(1 - \frac{326\pi^2}{3675}(1-3\kappa)\right)\Bigg]\end{aligned} \tag{9}$$

Here, $\zeta$ is the Reimann Zeta function, $t$ is a parameter given by $t = (\hbar c)(2k_B T d)^{-1}$, and $\delta$ and $\kappa$ are optical parameters given by:

$$\delta \equiv \frac{\delta_{Au} + \delta_{Ag}}{2} \tag{10}$$

$$\kappa \equiv \frac{\delta_{Au} + \delta_{Ag}}{(\delta_{Au} + \delta_{Ag})^2} \tag{11}$$

Where $\delta_{Au}$ and $\delta_{Ag}$ are the effective penetration depths of the electromagnetic oscillations into each metal film given by $\hbar c/\omega_p$ in which we have used $\omega_p$ = 9 eV for Au and $\omega_p$ = 8.6 eV for Ag.[36] Equation 9 is a limiting case of this theory in which $1/t \ll 1$, which is a valid approximation at our operating temperature ($T$ = 301.15 K) and $d < 1$ μm. Finally, a second order correction for surface roughness is included:[7,37]

$$F_C^R(d) = F_C^P(d)\left[1 + 6\left(\frac{A_r}{d}\right)^2\right] \tag{12}$$

Where $A_r$ is the stochastic RMS roughness amplitude of both surfaces. The surface roughness of the plate and sphere surfaces were found to be 2 nm and 8 nm respectively, using AFM (supplementary information S1). The total RMS roughness used in this model is $A_r = (A_{r,sphere}^2 + A_{r,plate}^2)^{1/2}$ = 8.25 nm.

Immediately following the electrostatic measurement, the plate is retracted back 1 μm away from the sphere and then stepped forward by 1 nm increments as $V_{bias}$ is adjusted according to the linear fit, ensuring that the first term of equation 1 is minimized at every position. At each plate position, 50,000 samples are taken in 0.5 seconds using the ADC. This measurement takes 20 minutes which requires that the enclosure temperature remain within 3 m°C to avoid unwanted thermal drift (see supplementary information S3).

After subtracting the zero-force signal ($c_3$), the measured data is scaled to units of force using the calibration factor $\gamma(d)$ measured from the electrostatic data. It is then fit to either the ideal Casimir theory (equation 8) or the corrected Casimir theory (equation 12) with $x_s$ as the only free parameter. These results can be seen in figure 3.

## Acknowledgments

The authors would like to thank Diego Perez, Jackson Chang, and Corey Pollock for helpful discussions, and David Lloyd for assistance with the collection of AFM data. This work is funded by the National Science Foundation grant no. 1708283, the Engineering Research Centers Program of the National Science Foundation through NSF

Cooperative Agreement under Grant EEC-0812056, Grant EEC-1647837, and Grant ECCS-1708283, and the DARPA Atoms to Product (A2P) Program/Air Force Research Laboratory (AFRL) contract no. FA8650-15-C-7545.

# Conflict of interests

All authors declare no competing financial interest.

# Author contributions

The device and experiments were conceived by D.B. and A.S. The fabrication was done by A.S. with assistance from J.J. All data was collected by A.S. and interpreted and analyzed by A.S., L.B., and D.B. The manuscript was written by A.S. with input from M.I. and edited by all authors.

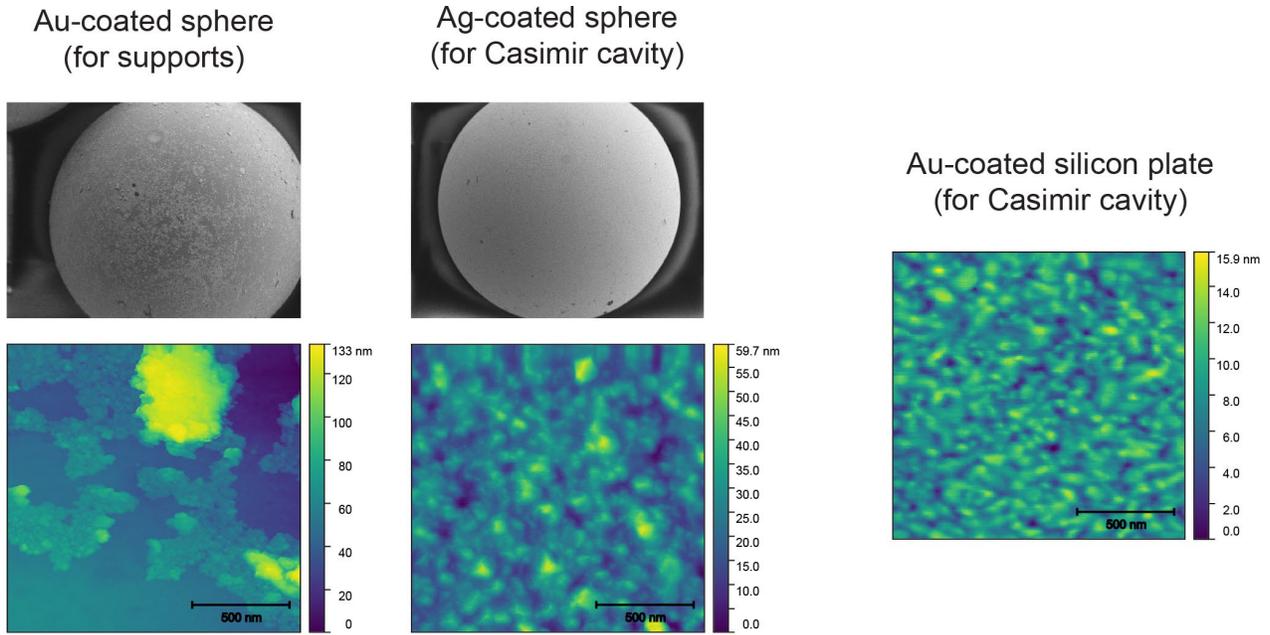

**S1:** SEM and AFM characterization of typical spherical and planar surfaces used in the modified accelerometer device. The gold spheres had abnormally distributed roughness and tended to contain large (> 1 μm) debris and were therefore only used for structural purposes. The silver spheres and gold coated silicon plate showed much lower RMS roughness and were used to construct the Casimir cavity. The data shown are taken of samples prepared in the same manner as those used in the experiment but are not scans of the specific surfaces in our Casimir cavity.

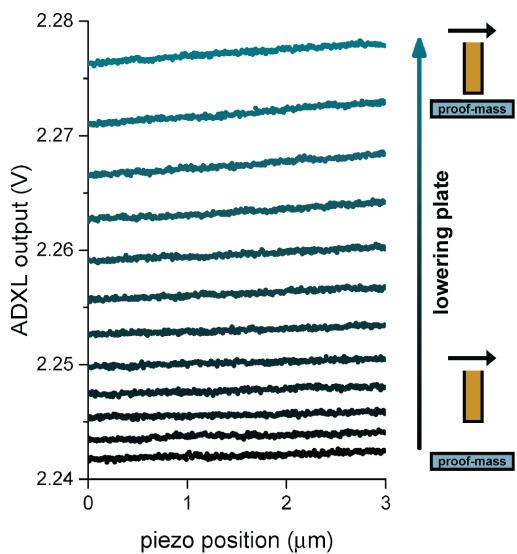

**S2:** Data illustrating the effect of the plate height on the accelerometer output due to interactions with the fringe feilds of the sensing electrodes. These scans were taken with the plate starting at approximately 12 μm above the proof-mass (black) and decreasing in height by 1 μm for each dataset. At each height, the plate is scanned by 3 μm in the X direction with the piezo stack actuator. Data was taken over a portion of the proof-mass far away from the sphere with the sphere and plate both grounded.

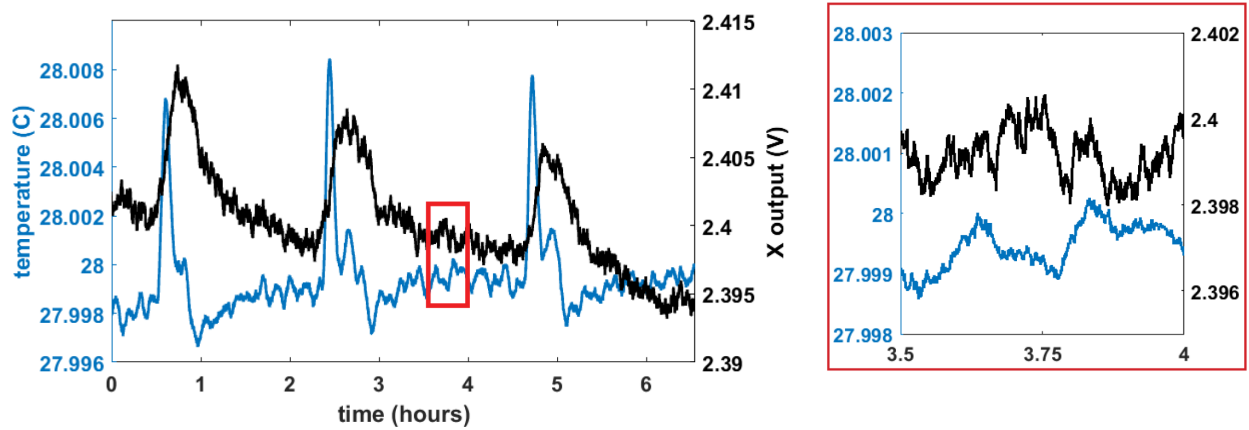

**S3:** Stability of the temperature and the accelerometer output in the temperature controlled enclosure over a 6.5 hour period. The periodicity of both datasets is a result from the heaters in the building turning on and off every two hours. Long term drift in the ADXL signal is also apparent. Between these unstable periods (red inset), our PID system is able to hold the temperature to within 3 m°C for 0.5 hours with is the length of a full experiment. For this data, the plate is placed 1 µm away from the sphere and a bias of 1 V is applied between the two surfaces resulting in an applied electrostatic force of approximately 1 nN.